\documentclass[12pt,a4paper,twoside]{article}
\usepackage{amssymb}
\usepackage{mathptmx}
\usepackage{theorem}
\usepackage{amsfonts,amsmath}

\newcommand{\R}{{\mathord{\mathbb R}}}
\newcommand{\Z}{{\mathord{\mathbb Z}}}
\newcommand{\N}{{\mathord{\mathbb N}}}
\newcommand{\C}{{\mathord{\mathbb C}}}

\newcommand{\T}{{\mathord{\mathbb T}}}

\newcommand{\mL}{{\mathcal L}}

\newcommand{\sign}{{\rm sign}}
\newcommand{\tr}{{\rm tr}}

\newcommand{\ch}{{\rm ch}}
\newcommand{\sh}{{\rm sh}}
\newcommand{\thh}{{\rm th}}
\newcommand{\Pf}{{\rm Pf}}

\begin{document}

\pagestyle{myheadings}

\markboth{W.H. Aschbacher, J.-M. Barbaroux}
{Out of equilibrium correlations in the XY chain}

\title{Out of equilibrium correlations in the XY chain}

\author{Walter H. Aschbacher$^{1}$\footnote{aschbacher@ma.tum.de}, \,Jean-Marie
Barbaroux$^{2}$\footnote{barbarou@univ-tln.fr}
\\ \\
$^1$Technische Universit\"at M\"unchen \\
Zentrum Mathematik, M5\\
85747 Garching, Germany
\\ \\
$^2$Centre de Physique Th\'eorique, Luminy\\
13288 Marseille, France\\
and D\'epartement de Math\'ematiques\\
Universit\'e du Sud Toulon-Var\\
83957 La Garde, France }

\maketitle


\begin{abstract}
We study the transversal spin-spin correlations in the non-equilibrium steady state of the XY chain constructed by
coupling a finite cutout of the chain to the two infinite parts to
its left and right acting as thermal reservoirs at different
temperatures. We prove that the spatial decay of these
correlations is at least exponentially fast.
\end{abstract}




\section{Introduction}

The XY chain  is the one-dimensional integrable spin system,
introduced in \cite{LSM61} (see also \cite{BR2}), whose formal
Hamiltonian is specified by
\begin{eqnarray}
\label{XY_Hamiltonian}
 H=-\frac{1}{4}
 \sum_{x\in\Z}\left\{(1+\gamma)\,\sigma_1^{(x)}\sigma_1^{(x+1)}
 +(1-\gamma)\,\sigma_2^{(x)}\sigma_2^{(x+1)}+2\lambda
 \, \sigma_3^{(x)}\right\},
\end{eqnarray}
where $\sigma_j^{(x)}$ denotes the Pauli matrix at site $x\in\Z$
in the transversal directions, $j=1,2$,  and in the longitudinal
direction, $j=3$. The parameter $\gamma\in (-1,1)$ describes the
anisotropy of the spin-spin coupling whereas $\lambda\in\R$ stands
for an external magnetic field.

Since the discovery of their ideal thermal conductivity,
quasi-one-dimensional Heisenberg  spin-$1/2$ systems, made from
different materials,  have been intensively studied experimentally
(see \cite{SFGOVR00, SGOVR01}; SrCuO$_2$ and Sr$_2$CuO$_3$ are
often considered to be the best physical
realizations of one-dimensional XYZ Heisenberg models).

Not only this  highly unusual transport property motivates the
theoretical study of such non-equilibrium models, but the XY chain
also represents one of the simplest non-trivial testing grounds
for the development of general ideas in rigorous non-equilibrium
theory.

Already in \cite{LSM61}, for vanishing external magnetic field
$\lambda=0$, the transversal spin-spin correlations in the ground
state and at nonzero temperature could be expressed by means of
determinants of large truncated Toeplitz matrices. Moreover, bounds on these
correlations implied vanishing long-range order, at least in the
isotropic case $\gamma=0$. Later, in \cite{MC68}, this study was
continued yielding an asymptotic evaluation of the transversal
correlations with the help of Szeg\"o's theorem. Afterwards, in
\cite{BMC71}, almost the complete phase diagram in $\gamma$ and
$\lambda$ for the behavior of the correlation functions both in the longitudinal and in the transversal directions were obtained, and so for zero and nonzero temperature: The result of this study for the case of nonzero temperature is that all the correlation functions are asymptotically exponentially decaying with a decay rate which depends on the magnetic field $\lambda$ and the anisotropy $\gamma$.

In this note, we study the large $n$ behavior of the transversal
spin-spin correlations,
\begin{eqnarray}
\label{Cj}
C_j(n)=\omega(\sigma_j^{(0)}\sigma_j^{(n)}),\quad j=1,2,\,\,n\in\N,
\end{eqnarray}
in the non-equilibrium steady state (NESS)
$\omega$ constructed in \cite{AP03}  (and already in \cite{AH} for $\gamma=0$) in a setting which has become
to serve as paradigm in non-equilibrium statistical mechanics: a
``small'' system which is coupled to two infinite reservoirs which
are in thermal equilibrium at different inverse temperatures
$\beta_L$ and $\beta_R$. We prove that the spatial decay of $C_j(n)$ is at least exponentially fast.  This behavior contrasts with the one generally expected by the folklore which predicts quasi-long-range order out of equilibrium.

In the following section, we give a brief informal description of our
non-equilibrium setting for the XY chain. We refer to
\cite{AJPP, AP03, JP02} for a precise formulation within the framework of
$C^\ast$ algebraic quantum statistical mechanics.

\section{The non-equilibrium setting for the XY chain}
\label{sec:non_eq_setting}

Consider the XY chain described by the Hamiltonian
\eqref{XY_Hamiltonian} and remove the two bonds at the sites $\pm
M$, $M>0$. Doing so, the  initial chain divides into a compound of
three noninteracting subsystems. This configuration is what we
call the free system whose Hamiltonian
$$
H_0=H_L+H_S+H_{_R}
$$
is built as in \eqref{XY_Hamiltonian} according to the union
$\Z_L\cup\Z_S\cup\Z_R$, where $\Z_L=\{x\in\Z\,|\,x\le-M-2\}$,
$\Z_R=\{x\in\Z\,|\,x\ge M+1\}$, and $\Z_S=\{x\in\Z\,|\,-M\le x\le
M-1\}$.  The infinite pieces $\Z_L$ and $\Z_R$ will play the role
of thermal reservoirs to which the finite system on  $\Z_S$ is
coupled by means of $V=H-H_0$. In contrast, the  configuration
described by $H$, i.e. the original XY
chain on the whole of $\Z$,  is considered to be the perturbed system.\\
In order to construct a NESS in the sense of \cite{R01}, we choose the
initial state $\omega_0$ to be composed of $(\tau_j,\beta_j)$-KMS
states $\omega_j$ on $\Z_j$, $j=L,R$, and of the
normalized trace state $\omega_S$ on $\Z_S$ ($\tau_j$
denotes the time evolution generated by $H_j$, and
$\beta_j$ is the inverse temperature), i.e. we set
$$
\omega_0=\omega_L\otimes \omega_S\otimes \omega_R.
$$
It is well-known that the Jordan-Wigner transformation maps the XY model
on a model of free fermions with annihilation and creation operators $c_x$, $c_x^\ast$. Using scattering theory on the 1-particle Hilbert space of these
fermions, the NESS $\omega$ w.r.t.  the initial state $\omega_0$ and the perturbed time evolution $\tau^t$ have been constructed in \cite{AP03},
$$
\omega(c_x)=\lim_{t\to\,+\infty}\omega_0(\tau^t(c_x)).
$$
It has been shown in \cite{AP03} that $\omega$ is a quasi-free
state with 2-point operator $S$,
\begin{eqnarray}
\label{ness}
\omega(B^\ast(f)B(g))=(f,Sg),
\end{eqnarray}
where $f,g\in l^2(\Z)\otimes\C^2\simeq l^2(\Z)\oplus l^2(\Z)$,
and $B(f)=\sum_{x\in\Z}(f_+(x)c_x^\ast+f_-(x)c_x)$ for
$f=(f_+,f_-)$. Moreover, $S$ has  been explicitly computed in
\cite{AP03}. In the Fourier picture, $l^2(\Z)\simeq L^2(\T)$ (with
$\T=\{z\in\C\,|\,|z|=1\}$),  it reads
\begin{eqnarray}
\label{Sorig}
S(\xi)=\left(1+e^{\beta \,h(\xi)+\delta \,k(\xi)}\right)^{-1},
\end{eqnarray}
with the parameters $\beta$ and $\delta$ given by
\begin{eqnarray}
\label{def:beta_delta}
\beta=\frac{1}{2}\left(\beta_R+\beta_L\right),\quad
\delta=\frac{1}{2}\left(\beta_R-\beta_L\right).
\end{eqnarray}
The $1$-particle operators $h$ and $k$ have the form
\begin{eqnarray*}
h(\xi)=(\cos\xi-\lambda)\otimes\sigma_3-\gamma\sin\xi\otimes\sigma_2,
\quad k(\xi)=\sign(\kappa(\xi))\,\mu(\xi)\otimes\sigma_0,
\end{eqnarray*}
where the  functions $\kappa(\xi)$ and $\mu(\xi)$ are defined by
\begin{eqnarray}
\label{kappa_mu} \kappa(\xi)=2\lambda\sin\xi-(1-\gamma^2)\sin
2\xi,\quad
\mu(\xi)=\left((\cos\xi-\lambda)^2+\gamma^2\sin^2\xi\right)^{1/2}.
\end{eqnarray}
Furthermore, $\sigma_0$ is the identity on $\C^2$, and $\sigma_1,
\sigma_2,\sigma_3$ denote the Pauli matrices,
\begin{eqnarray}
\label{PauliMatrices} \sigma_0=\left[\begin{array}{cc}1 & 0\\ 0&
1\end{array}\right], \quad \sigma_1=\left[\begin{array}{cc}0 & 1\\
1& 0\end{array}\right],\quad \sigma_2=\left[\begin{array}{cc}0 &
-i\\ i & 0\end{array}\right],\quad
\sigma_3=\left[\begin{array}{cc}1 & 0\\ 0& -1\end{array}\right].
\end{eqnarray}

\section{Exponential decay of the correlation function}
Let us now state our result.

\textbf{Theorem}\,\,{\em Let $\omega$ be the quasi-free NESS
characterized by \eqref{ness}, and $C_j(n)$, $j=1,2$,  the correlation
functions \eqref{Cj}. For all values of the parameters $0\leq\delta
<\beta < \infty$, $\gamma\in (-1,1)$, and $\lambda\in\R$, the transversal spin-spin correlations
$C_j(n)$ decay at least exponentially fast, and
 $$
  \limsup_{n\rightarrow\infty} \frac{\log \left|C_j(n)\right|}{n} \leq
  \frac12 \int_0^{2\pi} \!\frac{d\xi}{2\pi} \,\log\left[\thh (\beta_L\mu(\xi)/2)
  \,\thh(\beta_R\mu(\xi)/2)\right] <0.
 $$}

\vspace{0.2cm}

{\bf Remarks}\,\,{\em 1)  If the temperature difference of the reservoirs vanishes, i.e. $\delta=0$, the total system is in thermal equilibrium at inverse temperature $\beta_L=\beta_R$, and our bound estimates from above the asymptotically exponentially decaying behavior from \cite{BMC71}.

2) For $\delta\neq 0$, the correlation in the $3$-direction decays
like $1/n^2$ at infinity for all $\gamma\in(-1,1)$,
$\lambda\in\R$, see \cite{AP03}. In contrast to this result, our
theorem does not confirm the folklore about the change in the type
of decay -- from short range to long range -- when passing from
equilibrium to non-equilibrium.
}

\vspace{0.5cm}

Before we begin the proof of the theorem, we rewrite the
correlation function $C_j(n)$ as the determinant of a $2n\times
2n$ block Toeplitz matrix, in the same way as it has been done at
zero and non-zero temperature in \cite{BMC71, LSM61}. The reader
not familiar with Toeplitz theory may consult Appendix
\ref{app:Toeplitz} before entering the proof. We restrict our
attention to the transversal correlation function $C_1(n)\equiv
C(n)$ in the $1$-direction (the $2$-direction being analogous, see the Jordan-Wigner transformation for $\sigma_2$ in
\eqref{JordanWigner} below),
\begin{eqnarray*}
C(n)=\omega(\sigma_1^{(0)}\sigma_1^{(n)}),\quad n\in\N.
\end{eqnarray*}
The well-known Jordan-Wigner transformation expresses the spins
$\sigma_1^{(x)},\sigma_2^{(x)},\sigma_3^{(x)}$, $x\in\Z$,  by
means of fermionic creation and annihilation operators
$c_x^\sharp$ \,($=c_x^\ast,c_x$),
\begin{eqnarray}
\label{JordanWigner} \sigma_1^{(x)}=TS^{(x)}(c_x+c_x^\ast),\quad
\sigma_2^{(x)} =iTS^{(x)}(c_x-c_x^\ast),\quad \sigma_3^{(x)}
=2c_x^\ast c_x-1,
\end{eqnarray}
where $S^{(x)} = \sigma_3^{(1)} \ldots\sigma_3^{(x-1)}$ for $x>1$,
$S^{(1)}=1$, and $S^{(x)} = \sigma_3^{(x)} \ldots\sigma_3^{(0)}$
for $x<1$, see for example \cite{AP03}. The element $T$ stems
from Araki's $C^\ast$ crossed product extension of the CAR algebra
for the two-sided chain, see \cite{A84}. It has the properties
$T^2=1$, $Tc_0^\sharp=-c_0^\sharp T$, and $Tc_x^\sharp=c_x^\sharp
T$ for $x>0$. Plugging the Jordan-Wigner transformation
\eqref{JordanWigner} into the product
$\sigma_1^{(0)}\sigma_1^{(n)}$, we find
\begin{eqnarray*}
\sigma_1^{(0)}\sigma_1^{(n)} =(-1)^n
\,a^{(0)}_-\,a^{(1)}_+a^{(1)}_-\,a^{(2)}_+a^{(2)}_-\cdot...\cdot
a^{(n-1)}_+a^{(n-1)}_-\,a^{(n)}_+,
\end{eqnarray*}
where $a^{(x)}_\pm=c_x^\ast\pm c_x$. Moreover, we define
$\eta_\pm=(1,\pm 1)\in\C^2$ and $\alpha^{(x)}_\pm =
\delta_x\otimes \eta_\pm$ \,(with $\delta_x\in l^2(\Z)$,
$\delta_x(y)=1$ for $y=x$ and zero otherwise). With the linear
map $B$ introduced in \cite{A71}, $l^2(\Z)\oplus l^2(\Z)\ni
f=(f_+,f_-)\mapsto B(f)=\sum_{x\in\Z}(f_+(x)c_x^\ast+f_-(x)c_x)$,
we have  $a_\pm^{(n)}=B(\alpha_\pm^{(n)})$. Hence, we can express
the correlation function $C(n)$ as
\begin{eqnarray*}
C(n)=(-1)^n \,\omega(B(\alpha^{(0)}_-)\,B(\alpha^{(1)}_+)B(\alpha^{(1)}_-)\cdot...\cdot B(\alpha^{(n-1)}_+)B(\alpha^{(n-1)}_-)\,B(\alpha^{(n)}_+)).
\end{eqnarray*}
Since $\omega$ is quasi-free \cite{AP03}, the correlation function
$C(n)$ is a Pfaffian,
\begin{eqnarray*}
C(n)=\Pf\,\Omega(n),
\end{eqnarray*}
where the matrix $(\Omega(n)_{jk})_{j,k=1}^{2n}\in \C^{2n\times
2n}$  is defined to be skew-symmetric, and, for $j<k$,
\begin{eqnarray*}
\Omega(n)_{jk}=\omega(B(f_j)B(f_k))\quad\mbox{with}\quad f_{2i-1}=\alpha^{(i-1)}_-, \,\,f_{2i}=\alpha^{(i)}_+,\,\, i=1,2,...,n.
\end{eqnarray*}
The Pfaffian $\Pf\,A$ of a matrix $A\in \C^{2n\times 2n}$ is
defined  by
$\Pf\,A=\sum_{\pi}\sign(\pi)\prod_{k=1}^nA_{\pi_{2k-1},\pi_{2k}}$
with the sum running over all $\pi$ in the permutation group
$S_{2n}$ which satisfy $\pi_{2k},\pi_{2k+1}>\pi_{2k-1}$. If $A$ is
skew-symmetric, $A^T=-A$ ($A^T$ denotes the transpose of $A$), the
Pfaffian of $A$ is related to the determinant of $A$ through
$(\Pf\,A)^2=\det A$. Thus, we are led to study the large $n$
asymptotics of the determinant of $\Omega(n)$,
\begin{eqnarray}
\label{Cn2}
 C(n)^2=\det\Omega(n).
\end{eqnarray}
In order to cast $\Omega(n)$ into the form of a truncated block
Toeplitz matrix, we compute the matrix entries
\begin{eqnarray}
\label{A}
A^{jk}_{\pm\pm}=\omega(B(\alpha^{(j)}_\pm)B(\alpha^{(k)}_\pm)),\,\,1\le j<k\le n.
\end{eqnarray}
For this purpose, we rewrite the 2-point operator $S(\xi)$ on $L^2(\T)\otimes\C^2$ from \eqref{Sorig} as
\begin{eqnarray}
\label{S}
S(\xi)=s_0(\xi)\otimes \sigma_0+\sum_{k=1}^3s_k(\xi)\otimes \sigma_k,
\end{eqnarray}
using the Pauli matrices from \eqref{PauliMatrices}. The first
component $s_0(\xi)$ looks like
\begin{eqnarray}
\label{s0} s_0(\xi)&=&\frac{1}{2}+\frac{1}{2}
\,\sign(\kappa(\xi))\,\varphi_{\delta}(\xi),
\end{eqnarray}
and $s(\xi)=(s_1(\xi),s_2(\xi),s_3(\xi))$ has the form
\begin{eqnarray}
\label{s} s(\xi)&=&\frac{1}{2}\,\varphi_\beta(\xi)\,r(\xi),\quad
r(\xi)=\frac{1}{\mu(\xi)}\,(0,-\gamma \sin\xi,\cos\xi-\lambda),
\end{eqnarray}
with the functions $\kappa(\xi)$ and $\mu(\xi)$ from
\eqref{kappa_mu}.  Moreover, we used the definition
\begin{eqnarray}
\label{phi}
\varphi_\alpha(\xi)&=&\frac{\sh(\alpha \mu(\xi))}{\ch(\beta
 \mu(\xi))+\ch(\delta \mu(\xi))},\quad \alpha\in\R.
\end{eqnarray}
Let us start with the computation of $A^{jk}_{++}$ in \eqref{A}.
Using the property $B^\ast(f)=B(Jf)$ for $J: f=(f_+,f_-)\mapsto
({\bar f_-},{\bar f_+})$ from \cite{AP03}, and the expressions in
\eqref{S}, \eqref{s0}, and \eqref{s}, we find, for $1\le j<k\le
n$, and with the definition $e_j(\xi)=e^{-ij\xi}$, that
\begin{eqnarray*}
A^{jk}_{++}&=&(e_j\otimes\eta_+,S\,e_k\otimes\eta_+)_{L^2\otimes \C^2}\nonumber\\
&=&2 \int_0^{2\pi}\!\frac{d\xi}{2\pi}\,\,s_0(\xi) e^{-i(k-j)\xi}\nonumber\\
&=&\int_0^{2\pi}\!\frac{d\xi}{2\pi}\,\sign(\kappa(\xi))
\,\varphi_\delta(\xi)\,e^{-i(k-j)\xi}.
\end{eqnarray*}
The computation of $A^{jk}_{--}$ yields $A^{jk}_{--}=-A^{jk}_{++}$. Next, the entry $A^{jk}_{+-}$ looks like
\begin{eqnarray*}
A^{jk}_{+-}&=&(e_j\otimes\eta_+,S\,e_k\otimes\eta_-)_{L^2\otimes \C^2}\\
&=&2\int_0^{2\pi}\!\frac{d\xi}{2\pi}\,\,(s_3(\xi)+is_2(\xi))
e^{-i(k-j)\xi}\\
&=&\int_0^{2\pi}\!\frac{d\xi}{2\pi}\,\frac{\cos\xi-
\lambda-i\gamma\sin\xi}{\mu(\xi)}\,\,\varphi_\beta(\xi)\,e^{-i(k-j)\xi}.
\end{eqnarray*}
Finally, $A^{jk}_{-+}=-A^{kj}_{+-}$. Due to translation
invariance, we can define $A^{x}_{\pm\pm}=A^{jj+x}_{\pm\pm}$ with
$x\in\Z$. Using $A_{--}^x=-A_{++}^x$, \,$A_{-+}^x=-A_{+-}^{-x}$,
and the symmetry property $A^{-x}_{++}=-A^{x}_{++}$, we can write
$\Omega(n)$ in \eqref{Cn2} in the form of a $2n\times 2n$
truncated block Toeplitz matrix with $2\times 2$ blocks $a_x$,
\begin{eqnarray}
\label{Omega_n}
\Omega(n)=\left[
    \begin{array}{cccc}
    a_0 & a_{-1} & \ldots & a_{-(n-1)}\\
    a_1 & a_0    &  \ldots & a_{-(n-2)}\\
   \vdots & \vdots & \ddots & \vdots \\
    a_{n-1} & a_{n-2} & \ldots & a_0 \\
    \end{array}\right],
\end{eqnarray}
where the blocks are given by
\begin{eqnarray}
\label{def:FourierCoeff}
a_{x}=\left[\begin{array}{ll} A_{++}^{x} & -A_{+-}^{x-1} \\ A_{+-}^{-x-1} &
-A_{++}^{x}\end{array}\right],\quad x\in\Z.
\end{eqnarray}

Using the relations $a_x=\int_0^{2\pi}d\xi/(2\pi)
\,a(\xi)e^{-ix\xi}$ from \eqref{def:symbol}, we can extract from
\eqref{def:FourierCoeff}  the symbol $a(\xi)$
\begin{eqnarray}
\label{ssymbol} a(\xi)=\left[
    \begin{array}{ll}
    \sign(\kappa(\xi))\,\varphi_\delta (\xi) & -q(\xi)
    \,\varphi_\beta (\xi) \\
      {\bar q}(\xi)\,\varphi_\beta (\xi) & -\sign(\kappa(\xi))
      \,\varphi_\delta(\xi)
    \end{array}
\right].
\end{eqnarray}
Here,
$\varphi_\alpha(\xi)$ is defined in \eqref{phi}, $\kappa(\xi)$,
$\mu(\xi)$ in \eqref{kappa_mu}, and ${\bar q}(\xi)$ is the complex conjugate of
\begin{eqnarray*}
q(\xi)=\frac{\cos\xi-\lambda-i\gamma\sin\xi}{\mu(\xi)}\,e^{i\xi}.
\end{eqnarray*}

Therefore, denoting by $T[a]$ the block Toeplitz matrix associated
with the symbol $a$ given by \eqref{ssymbol}, we can write $\Omega(n)$ as a truncated block Toeplitz matrix (see \eqref{truncated}),
 $$
  \Omega(n)=T_n[a]= P_nT[a]P_n.
 $$

\vspace{0.5cm}

{\bf Proof of the Theorem}\,\,For $n\in\N$, let $s_1^{(n)}\leq s_2^{(n)} \leq \ldots \leq
s_{2n}^{(n)}$ be the  singular values of $T_n[a]$,
and denote by $N_{\epsilon,n}$ the number of singular values in
$[0,\epsilon]$, $0<\epsilon<1$, including multiplicity. From the above computations, we have
\begin{equation*}
  |C(n)|^2  = |\det \Omega(n)|= |\det T_n[a]| = \prod_{j=1}^{2n} s_j^{(n)}\leq \epsilon^{N_{\epsilon,n}} \prod_{j = N_{\epsilon,n}+1}^{2n}
 s_j^{(n)}
\end{equation*}
Thus,
\begin{eqnarray*}
  \log |\det T_n[a]| & \leq & \sum_{j = N_{\epsilon,n}+1}^{2n}
  \log s_j^{(n)}  \leq \sum_{j = 1}^{2n}
  \chi_\epsilon (s_j^{(n)}) \log s_j^{(n)},
\end{eqnarray*}
where $0\leq\chi_\epsilon\leq 1$ is a smooth characteristic
function with support in $(\epsilon,\|T[a]\|+1)$ and
value $1$ on $[\epsilon + \epsilon^2, \|T[a]\|]$. Since the symbol $a$ of $T_n[a]$ has the two singular values $\varphi_\beta - \varphi_\delta$ and $\varphi_\beta +
\varphi_\delta$, the Avram-Parter theorem \cite[p.203]{BS99} (see also \eqref{AvramParter}) yields
\begin{eqnarray*}
 \limsup_{n\rightarrow\infty}  \frac1n\log |\det T_n[a]|
 & \leq & \int_{0}^{2\pi}\!\frac{d\xi}{2\pi}\,
 \tr \left[(\chi_{\epsilon}\log )(|a(\xi)|)\right] \\
 & = & \int_{0}^{2\pi}\! \frac{d\xi}{2\pi}\,
   (\chi_{\epsilon}\log)(\varphi_\beta(\xi)-\varphi_\delta(\xi))
  +(\chi_{\epsilon}\log)(\varphi_\beta(\xi)+\varphi_\delta(\xi)) .
\end{eqnarray*}
Since the inequality holds for all $\epsilon>0$, and since for all
$\xi\in [0,2\pi]$ we have $0<\varphi_\beta(\xi) -
\varphi_\delta(\xi) \leq \varphi_\beta(\xi) +
\varphi_\delta(\xi)<1$, we obtain
\begin{eqnarray}
\label{bound}
 \limsup_{n\rightarrow\infty}  \frac1n\log |\det T_n[a]|
 & \leq & \int_0^{2\pi}\! \frac{d\xi}{2\pi}\,\log\left[\thh(\beta_L\mu(\xi)/2)
 \,\thh(\beta_R\mu(\xi)/2)\right]
 \end{eqnarray}
 which concludes the proof. \hfill $\Box$

\vspace{1cm}

{\bf Remarks}\,\, {\it 3) We do not provide a lower bound in our proof (but see remark 1 for $\delta=0$). Due to the lack of results on the invertibility of block Toeplitz matrices $T[a]$,  the behavior of the singular values of $T_n[a]$ close to zero is, in general, rather difficult to control.

4) In the critical regime of the XY chain, $\gamma=0$, $\lambda\in [-1,1]$, and $\gamma\neq 0$, $\lambda=\pm 1$, where the nonnegative function $\mu(\xi)$ from \eqref{kappa_mu} can have zeroes on $[0,2\pi]$, the bound \eqref{bound} remains finite due to the integrability of $\log x$ at the origin.

5) A straightforward proof using the inequalities $|\det
T_n[a]|\leq \|T_n[a]\|^{2n}$ and}
$$
\|T_n[a]\| \leq \|T[a]\| =\mbox{ess} \sup_{\xi\in
\T}\,\|a(\xi)\|_{\mL(\C^N)}=\mbox{ess} \sup_{\xi\in
\T}\,s_2(a(\xi))\quad
$$
{\it (see \cite{BS99} and \eqref{ToeplitzNorm}-\eqref{def:phi_infinity}) yields the weaker bound $|\det T_n[a]|\le
\thh^{2n}(\beta_R\|\mu\|_\infty/2)$ which, in contrast to \eqref{bound}, is valid for all $n$. On the other hand, in the limit $\beta_R\to\infty$, this bound is void whereas \eqref{bound} still implies exponential decay as long as $\beta_L<\infty$.
}

\appendix
\section{Toeplitz operators}
\label{app:Toeplitz}

{\bf Toeplitz operators} \cite[p.185]{BS99}\,\,  Let $N\in\N$. We
define the space $l^2_N$ of all $\C^N$-valued sequences
$f=\{f_i\}_{i=1}^\infty$, $f_i\in\C^N$, by
\begin{eqnarray}
\label{l2N} l^2_N=\{f:\N\to\C^N\,|\, \|f\|<\infty\},\quad
\|f\|=\left(\sum_{i=1}^\infty\|f_i\|^2_{\C^N}\right)^{1/2},
\end{eqnarray}
where $\|\cdot \|_{\C^N}$ denotes the $l^2$ norm on $\C^N$. We
write $l^2\equiv l^2_1$. Let $\{a_x\}_{x\in\Z}$ be a sequence of
$N\times N$ matrices, $a_x\in\C^{N\times N}$. The Toeplitz
operator defined through its action on elements of $l^2_N$ by
$f\mapsto \{\sum_{j=1}^\infty a_{i-j} \,f_j\}_{i=1}^\infty$ is a
bounded operator on $l^2_N$, if and only if
\begin{eqnarray}
\label{def:symbol}
a_x=\int_0^{2\pi}\frac{d\xi}{2\pi}\,a(\xi) e^{-ix\xi}
\end{eqnarray}
for some $a\in L^\infty_{N\times N}$ (see \cite[p.186]{BS99}),
where we define (with $\T=\{z\in\C\,|\,|z|=1\}$)
\begin{eqnarray}
\label{def:LNN}
L^\infty_{N\times N}=\{\phi:\T\to\C^{N\times N}\,|\,
\phi_{ij}\in L^\infty(\T), i,j=1,...,N\}.
\end{eqnarray}
In this case, we write the Toeplitz operator as
\begin{eqnarray}
\label{def:Toeplitz}
T[a]=\left[
    \begin{array}{lllll}
    a_0 & a_{-1} & a_{-2} & ...\\
    a_1 & a_0 & a_{-1} & ...\\
    a_2 & a_1 & a_0 &... \\
    ... & ... & ... & ...\\
    \end{array}\right].
\end{eqnarray}
The function $a\in L^\infty_{N\times N}$ is called the symbol of
$T[a]$. If $N=1$ the symbol $a\in L^\infty\equiv L^\infty_{1\times
1}$ and the Toeplitz operator $T[a]$ are called scalar, whereas
for $N>1$ they are called block.

For $n\in\N$, let $P_n$ be the projections on $l^2_N$,
\begin{eqnarray*}
 P_n(\{x_1,...,x_n,x_{n+1},...\})=\{x_1,...,x_n,0,0,...\}.
\end{eqnarray*}
With the help of these $P_n$, we define the truncated $Nn\times Nn$ Toeplitz matrices as
\begin{equation}\label{truncated}
 T_n[a]= P_n T[a] P_n|_{\mathrm{Im\,}P_n} .
\end{equation}

{\bf Norm of Toeplitz operators} \cite[p.186]{BS99}\,\, The norm
of a Toeplitz operator is related to its symbol,
\begin{eqnarray}
\label{ToeplitzNorm}
\|T[a]\|=\|a\|_\infty,
\end{eqnarray}
where $\|a\|_\infty$ is defined to be the operator norm of the
multiplication operator acting on $\oplus_{1}^NL^2(\T)$ by
multiplication with the matrix function $a\in L^\infty_{N\times
N}$. From \cite[p. 95]{BS90}, we have
\begin{eqnarray}
\label{def:phi_infinity}
\|a\|_\infty=\mbox{ess} \sup_{\hspace{-0.5cm}\xi\in
\T}\,\|a(\xi)\|_{\mL(\C^N)},
\end{eqnarray}
where $\|\cdot\|_{\mL(\C^N)}$ is the operator norm induced by the $l^2$ norm on $\C^N$.

{\bf Avram-Parter theorem} \cite[p.203]{BS99}\,\, Let
$s_1^{(n)},...,s_{Nn}^{(n)}$ be the singular values of the
truncated Toeplitz operator $T_n[a]$ with block symbol $a\in
L^\infty_{N\times N}$, and $g$ a continuous function with compact
support, $g\in C_0(\R)$. Then,
\begin{eqnarray}
\label{AvramParter}
\lim_{n\to\infty}\frac{1}{Nn}\sum_{j=1}^{Nn}g(s_j^{(n)})
=\frac{1}{N}\int_0^{2\pi}\!\frac{d\xi}{2\pi}\,\tr \,g(|a(\xi)|),
\end{eqnarray}
where $\tr A$ is the trace of the matrix $A\in \C^{m\times m}$,
$m\in\N$, and $|A|=\sqrt{A^*A}$ its modulus.


\vspace{2cm}

\textbf{Acknowledgements}\quad We gratefully acknowledge
the financial support by {\bf ACI} {\it "Mod\'elisation
stochastique des syst\`emes hors \'equilibre"}, Minist\`ere
d\'el\'egu\'e \`a la Recherche, France.\\
Moreover, we would like to thank the referee for his constructive remarks.



\end{document}